\documentclass[
preprint,
floatfix,
pra,
showpacs
]{revtex4}

\usepackage{mathrsfs}
\usepackage{braket}
\usepackage{graphicx}
\graphicspath{{.},{pics/}}

\begin{document}


\title{Coherent state LOQC gates using simplified diagonal superposition resource states}


\author{A. P. Lund}
\email{lund@physics.uq.edu.au}
\author{T. C. Ralph}
\affiliation{Centre for Quantum Computer Technology, Department of Physics,\\
University of Queensland, St Lucia, QLD 4072, Australia}


\pacs{03.67.Lx, 42.50.Dv}

\begin{abstract}

In this paper we explore the possibility of fundamental tests for coherent state optical quantum computing gates [T.~C.~Ralph, et. al, Phys. Rev. A \textbf{68}, 042319 (2003)] using sophisticated but not unrealistic quantum states.  The major resource required in these gates are state diagonal to the basis states.  We use the recent observation that a squeezed single photon state ($\hat{S}(r) \ket{1}$) approximates well an odd superposition of coherent states ($\ket{\alpha} - \ket{-\alpha}$) to address the diagonal resource problem.  The approximation only holds for relatively small $\alpha$ and hence these gates cannot be used in a scaleable scheme.  We explore the effects on fidelities and probabilities in teleportation and a rotated Hadamard gate.  
\end{abstract}


\maketitle



\section{Introduction}

It was long believed that optical quantum computing would require enormous non-linear interactions between optical modes in order to be a viable technology.  This was mainly due to the requirement that the presence of a single photon in an optical mode must control the path of another photon (see for example~\cite{milburn:fredkingate}).  However, it has been shown that linear interactions combined with post-selective measurements induce enough non-linearity so that in principle one can perform quantum computation efficiently~\cite{KLM}.  The fundamental gates in this scheme work non-deterministically, but this can be overcome by quantum gate teleportation~\cite{nielsen:gateteleportation}.  To achieve near deterministic teleportation by linear interactions and post-selection requires a large linear network involving many modes prepared in single photon states~\cite{KLM}.  

More recently an alternative proposal which uses two coherent states and superpositions thereof (i.e. cat-like states) for quantum computing has emerged~\cite{ralph:catcomputing}.  Provided the two coherent states are sufficiently well seperated in phase space, the fundamental gates of this scheme are near deterministic.  The gates described in this scheme consume equal superpositions of coherent states as a resource.  Generation of such states at the large separations required is a formidable challenge.  However, as has been recently reported~\cite{lund:catgeneration}, superpositions of coherent states that are not so well separated are well approximated by squeezed single photon states.  Here, we explore the possibility of constructing some of the gates outlined in~\cite{ralph:catcomputing} in this intermediate regime using the squeezed single photon as the superposition state resource.  We find it is possible to see the desired effects with fairly high (though not unit) visibility.

\section{The squeezed single photon as a superposition of coherent states}

It is shown in~\cite{ralph:catcomputing} that one can construct a universal set of gates used for quantum computing encoding a two level system in coherent states $\ket{\alpha}$ and $\ket{-\alpha}$ provided $\alpha$ is large enough (i.e. $\alpha \approx 2$).  One requirement for constructing these gates is a source of states which are diagonal superpositions of the basis states.  That is states of the form
\begin{equation}
\ket{\alpha} \pm \ket{-\alpha}.
\end{equation}  
Following~\cite{ralph:catcomputing} we will call these states ``cat states''.  The state with a plus sign is of even parity so we will call it an even cat state and the minus sign is of odd party so we will call it an odd cat state.  The coherent amplitude $\alpha$ will sometimes be referred to as the size of the cat state.  

\begin{figure}
\includegraphics[width=7.5cm]{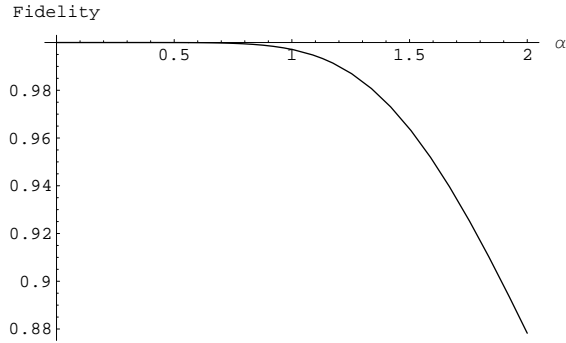}
\caption{The maximum possible fidelity obtained using the squeezed single photon as a source of odd cat states with a given size $\alpha$.  This given size is varied along the x-axis.}
\label{sqzphoton-fidelity}
\end{figure}

A recent observation is that the odd cat state is well approximated by a ``squeezed'' single photon state~\cite{lund:catgeneration}.  
The squeezed single photon state is of a simple analytic form so that the state can be written down and quantities of interest can be calculated exactly.  In terms of a vacuum state and annihilation and creation operators the state can be written
\begin{equation}
\hat{S}(r) \ket{1} = \hat{S}(r) \hat{a}^\dagger \ket{0}
\end{equation}
where $\hat{S}(r) = e^{\frac{r}{2}(\hat{a}^2 - \hat{a}^{\dagger 2})}$ called the ``single mode squeezing'' operator or just the ``squeezing operator''.  Here $r$ is a real parameter.  This operator reduces the noise seen in a quadrature measurement of the oscillator in the vacuum state by a factor of $e^{-r}$.
Because $r$ is assumed real it is possible to show (see~\cite{walls:qo})
\begin{equation}
\hat{a}^\dagger \hat{S}(r) = \hat{S}(r) \hat{a}^\dagger  \cosh r + \hat{S}(r) \hat{a}  \sinh r
\end{equation}
and hence
\begin{equation}
\label{sqz_ident}
\hat{a}^\dagger \hat{S}(r) \ket{0} =  \cosh r \hat{S}(r) \hat{a}^\dagger \ket{0}.
\end{equation}
The squeezing operator applied to the vacuum state generates the squeezed vacuum states.  They can be expanded in terms of photon number states as
\begin{equation}
\label{sqz_vac}
\hat{S}(r) \ket{0} = \sum_{n=0}^\infty \frac{(-\tanh r)^n}{\sqrt{\cosh r}} \frac{(2n!)^\frac{1}{2}}{2^n n!} \ket{2n}.
\end{equation}
Using equations~\ref{sqz_ident} and~\ref{sqz_vac}  one obtains the expansion for a squeezed single photon
\begin{equation}
\label{sqz_pho}
\hat{S}(r) \ket{1} = \sum_{n=0}^\infty \frac{(-\tanh r)^n}{(\cosh r)^{\frac{3}{2}}} \frac{\sqrt{(2n + 1)!}}{2^n n!} \ket{2n + 1}.
\end{equation}
The ``fidelity'' is a measure of how close two states are.  Fortunately all reference states that we wish to compare other states to will be pure states.  So here we call $\bra{\psi} \hat{\rho} \ket{\psi}$ the fidelity where $\ket{\psi}$ is the desired pure state and $\hat{\rho}$ is the density operator of the state actually generated.
Computing the fidelity of the state in equation~\ref{sqz_pho} with that of an odd cat state with size $\alpha$ one obtains
\begin{equation}
\label{fidelity}
\mathscr{F} (\alpha, r) = \frac{e^{-\alpha^2}}{2(1-e^{-2\alpha^2})} \frac{4 \alpha^2}{(\cosh r)^3} e^{-\alpha^2 \tanh r}.
\end{equation}
If one wishes to produce an odd cat state of size $\alpha$ (i.e. $\alpha$ is a given constant) then the fidelity is maximised when $r$ satisfies
\begin{equation}
\label{sqz-par}
r = \textrm{arccosh} \left( \sqrt{\frac{1}{2} + \frac{1}{6} \sqrt{9 + 4 \alpha^2}} \right).
\end{equation}
Substituting this relationship into the equation~\ref{fidelity} reduces it to a function for fidelity which depends on $\alpha$ alone.  This is the highest possible fidelity for a cat state of size $\alpha$ given that it was produced using the squeezed single photon state.  This function is plotted for $\alpha \in [0,2]$ in figure~\ref{sqzphoton-fidelity}. 
The high fidelity for $\alpha$ small is due to the odd cat state being dominated by its lowest photon number state; i.e. the odd cat state for $\alpha$ very small contains only a single photon.  When $\alpha = 0$ is substituted into equation~\ref{sqz-par}, the result is $r = 0$.  Hence no squeezing is performed and the state is just a single photon.  These two states are identical giving unit fidelity.  The fidelity remains high for $\alpha$ small as the next dominant term in the odd cat state for small $\alpha$ is the three photon term.  The squeezing of the single photon will coherently add pairs of photons to the single photon state.  Hence the ratio of the one and three photon state coefficients can be matched well by adjusting the squeezing level, provided that the next term (the 5 photon term) remains small.  Eventually as $\alpha$ increases these higher terms cannot be matched and the fidelity falls.  As $\alpha \rightarrow \infty$ the fidelity tends towards zero.   

\section{Coherent state quantum computing (CSQC) and teleportation}

\subsection{Coherent state quantum computing}

As stated above one may consider the states $\ket{\alpha}$ and $\ket{-\alpha}$ to be a basis for a two level quantum system.  If the two states are sufficiently distinguishable (i.e. $\alpha > 2$) one may also consider them to be an orthonormal basis for a two level system~\cite{ralph:catcomputing} to a very good approximation.  This two level quantum system is suitable for encoding quantum binary digits (qubits).  The phase of the coherent amplitude (i.e. the plus or minus sign) is utilised to encode information.  It is shown in~\cite{ralph:catcomputing} how one can build near-deterministic gates to perform universal quantum computation using these states as qubits.  We will call the procedures described in~\cite{ralph:catcomputing} used for universal quantum computing collectively as Coherent State Quantum Computing (CSQC).  

The one main resource that CSQC requires is a source of states diagonal to the basis states (i.e. the even or odd cat state).  One procedure which is crucial to gate operation is the ability to perform teleportation on qubits in this encoding.  This can be performed by using an odd cat state as shown later in this section.

Generating states diagonal to the coherent state basis with large coherent amplitude in a propagating optical mode is a formidable challenge.  However, as we have observed the squeezed single photon is a good approximation to the diagonal states provided $\alpha$ is not too large.  Generation of squeezed single photon states seems more experimentally accessible than alternative proposals (see~\cite{lund:catgeneration} in the short term motivating us to consider if in principle demonstrations of basic gate operations are possible using squeezed single photons as our resource state.  For example, consider $\alpha = 1$.  From figure~\ref{sqzphoton-fidelity} we observe that $\hat{S}(r)\ket{1}$ is still an excellent approximation to a cat state of this size whilst the overlap between $\ket{\alpha}$ and $\ket{-\alpha}$ has already fallen to $|\braket{-\alpha | \alpha}|^2 \approx 0.02$.  This suggests that interesting tests of principle can be carried out in this ``middle-ground''.  

\subsection{Teleportation of coherent state qubits}

The most basic gate in the CSQC scheme (after a $\hat{X}$ gate which is simply a phase shift of $\alpha$) is the teleportation gate.  This gate is also crucial in implementing a $\hat{Z}$ gate and required ``projections'' onto the spaced spanned by the two states $\ket{\alpha}$ and $\ket{-\alpha}$.  So let us consider the properties of this gate in more detail.  (Initially we will be considering exact superpositions of coherent states and not the squeezed single photon approximation).    

\subsubsection{CSQC Bell state generation and Bell state measurements}

In order to perform teleportation one must be able to create a Bell state and perform a measurement in the Bell basis~\cite{bennett:teleportation}.  Following~\cite{jeong:catbellstates} when two modes of the EM field are combined at an asymmetric 50:50 beam-splitter, the action writen in terms of the bell states using the encoding above is
\begin{eqnarray}
\ket{\alpha, \alpha} + \ket{-\alpha, - \alpha} & \rightarrow & \ket{0} \otimes \left( \ket{\sqrt{2} \alpha} + \ket{-\sqrt{2} \alpha} \right) \\
\ket{\alpha, \alpha} - \ket{-\alpha, - \alpha} & \rightarrow & \ket{0} \otimes \left( \ket{\sqrt{2} \alpha} - \ket{-\sqrt{2} \alpha} \right) \label{cat2} \\
\ket{\alpha, -\alpha} + \ket{-\alpha, \alpha} & \rightarrow &  \left( \ket{\sqrt{2} \alpha} + \ket{-\sqrt{2} \alpha} \right) \otimes \ket{0}  \\
\ket{\alpha, -\alpha} - \ket{-\alpha, \alpha} & \rightarrow &  \left( \ket{\sqrt{2} \alpha} - \ket{-\sqrt{2} \alpha} \right) \otimes \ket{0}
\end{eqnarray}
where the notation $\ket{\alpha, \beta} \equiv \ket{\alpha} \otimes \ket{\beta}$ has been used and normalisation factors have been ignored.  This transformation follows from the linear evolution of quantum states and the expected addition and subtraction of coherent state amplitudes at a beamsplitter.  So the four Bell states can be distinguished by measuring one mode to be the vacuum and then determining if the other mode contains an odd or even state.  For example the first state is chosen if `zero' is measured in the first mode and an even number in the second due to the even cat state present in this mode.  The second state is selected when counting `zero' and `odd', the third state `even' and `zero' and the fourth state `odd' and `zero' count pairs.  Note that an \emph{even number of photons includes zero}.  This means that a `zero' and `zero' measurement pair can occur for the first and third states leaving them undistinguished.  So if `even' excludes the possibility of zero, then the states can be distinguished but when a `zero' and `zero' measurement occurs the measurement has failed.  This is a consequence of the non-orthogonality of the qubit encoding.  When not working in the range of the squeezed single photon approximation this `zero' and `zero' possibility can be made arbitrarily smaller by making $\alpha$ large.  

\begin{figure}
\setlength{\unitlength}{1.5cm}
\centerline{
\begin{picture}(3.5,2.75)
\put(0,0){\line(1,0){0.5}}
\put(0,1){\line(1,0){0.5}}
\put(0,2){\line(1,0){1.5}}
\put(0.5,0){\line(1,1){2}}
\put(0.5,1){\line(1,-1){1}}
\put(1.5,0){\vector(1,0){2}}
\put(1.5,2){\line(1,-1){1}}
\put(2.5,1){\line(1,0){0.5}}
\put(2.5,2){\line(1,0){0.5}}
\put(0.75,0.5){\line(1,0){0.5}}
\put(1.75,1.5){\line(1,0){0.5}}
\put(0,0.1){\makebox(0.5,0.5)[bl]{$\ket{0}$}}
\put(-1,1.1){\makebox(0.5,0.5)[bl]{$\ket{\sqrt{2}\alpha}-\ket{-\sqrt{2}\alpha}$}}
\put(-1,2.1){\makebox(0.5,0.5)[bl]{$\mu \ket{\alpha}+\nu\ket{-\alpha}$}}
\put(2.75,2.1){\makebox(0.5,0.5)[bl]{Count ``0''}}
\put(2.75,1.1){\makebox(0.5,0.5)[bl]{Count ``odd''}}
\put(2.75,0.1){\makebox(0.5,0.5)[bl]{Output}}
\put(1.3,0.25){\makebox(0.5,0.5)[l]{0.5}}
\put(2.3,1.25){\makebox(0.5,0.5)[l]{0.5}}
\end{picture}
}
\caption{A schematic diagram of the teleporter.  The lower two modes after the first beamsplitter contain the entangled pair.  The top mode contains the qubit.  The bell state measurement is made on one half of the entanglement and the qubit by the second beamplitter.  Only one of the bell state measurements is accepted here by the zero, odd count.  When this occurs the lower mode contains the input qubit without any corrections needed.}
\label{teleport}
\end{figure}
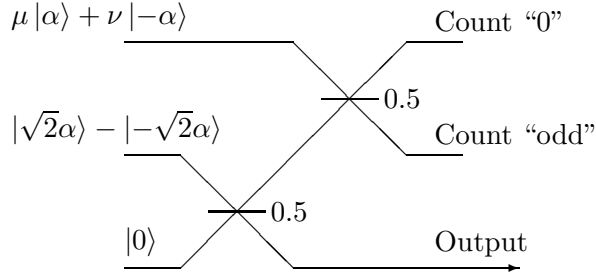

To perform teleportation one requires a prepared state in one of the four Bell states.  If one reverses the procedure of the Bell state measurement then it can be seen that this takes a non-entangled state to an entangled state.  With the usage of the squeezed single photon state in mind, one could use this state and the inverse of equation~\ref{cat2} to create the entangled Bell state.  A Bell state measurement is then performed on the input qubit and one half of this entangled state as just described.  A schematic diagram of this configuration is shown in figure~\ref{teleport}.  

As an example of how the entire state evolves during this process we can write out the composite system of an arbitrary qubit and the Bell state from equation~\ref{cat2} as
\begin{widetext}
\begin{equation}
\label{ent_sys}
\mu \left( \ket{\alpha, \alpha, \alpha} - \ket{\alpha,-\alpha,-\alpha} \right) + \nu \left( \ket{-\alpha,\alpha,\alpha} - \ket{-\alpha,-\alpha,-\alpha} \right).
\end{equation}
\end{widetext}
Here the notation introduced above has been used to combine modes and the entanglement is present in the second and third modes.  The modes in this state correspond to the top (first label), middle (second label) and bottom (third label) modes in figure~\ref{teleport}.  The teleportation procedure now requires a bell basis measurement on the qubit and one half of the entangled pair.  As explained this is done by applying a 50:50 beamsplitter on the first two modes:
\begin{widetext}
\begin{equation}
\mu \left( \ket{0, \sqrt{2} \alpha,\alpha} - \ket{\sqrt{2}\alpha,0,-\alpha} \right) + \nu \left( \ket{-\sqrt{2}\alpha, 0, \alpha} - \ket{0, -\sqrt{2}\alpha,-\alpha} \right).
\end{equation}
\end{widetext}
Then if the Bell state in equation~\ref{cat2} is projected on to by performing a $\ket{0,odd}$ number measurement, the state in the third mode is
\begin{equation}
\mu \ket{\alpha} + \nu \ket{-\alpha}
\end{equation}
and successful teleportation has occurred.  Note that to distinguish the $\ket{0,odd}$ states from the $\ket{0,even}$ states requires very efficient photon number resolving measurements.  The loss of a single photon will change the odd result to an even result.

\subsubsection{Corrections and probability of success}

The other Bell basis measurement events can be used boosting the overall probability of success.  However $\hat{X}$ (bit flip) and $\hat{Z}$ (phase) corrections must be applied to the output depending on which result was obtained.  The $\hat{X}$ correction can be applied by applying a $\pi$ phase shift to the output mode.  The $\hat{Z}$ correction is more difficult and needs be applied when an even cat state is detected in the Bell state analysis.  One possible solution proposed in~\cite{ralph:catcomputing} is to apply teleportation again in the hope that another $\hat{Z}$ correction is required cancelling out the $\hat{Z}$ applied in the initial teleportation. 

\begin{figure}
(a)\includegraphics[width=0.3\textwidth]{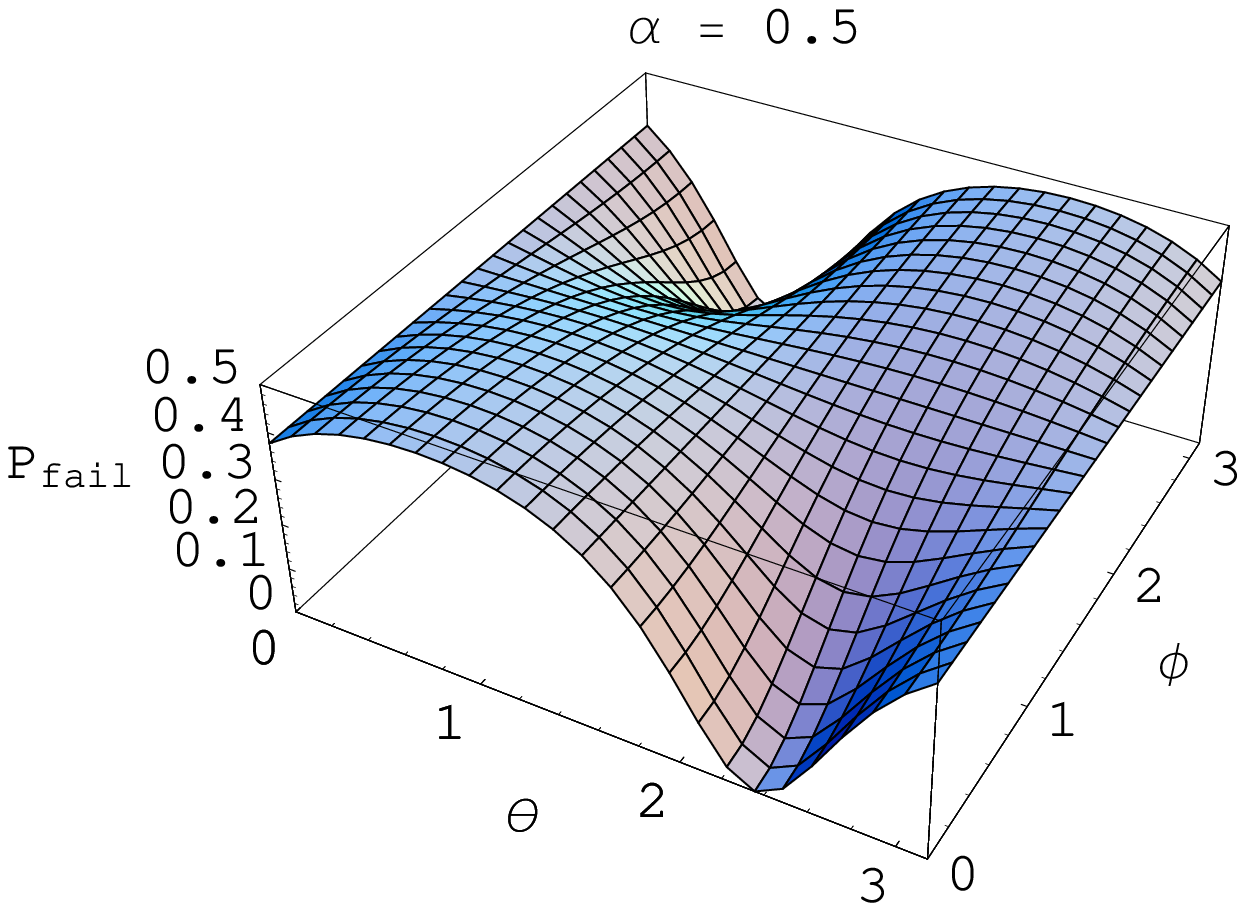}
(b)\includegraphics[width=0.3\textwidth]{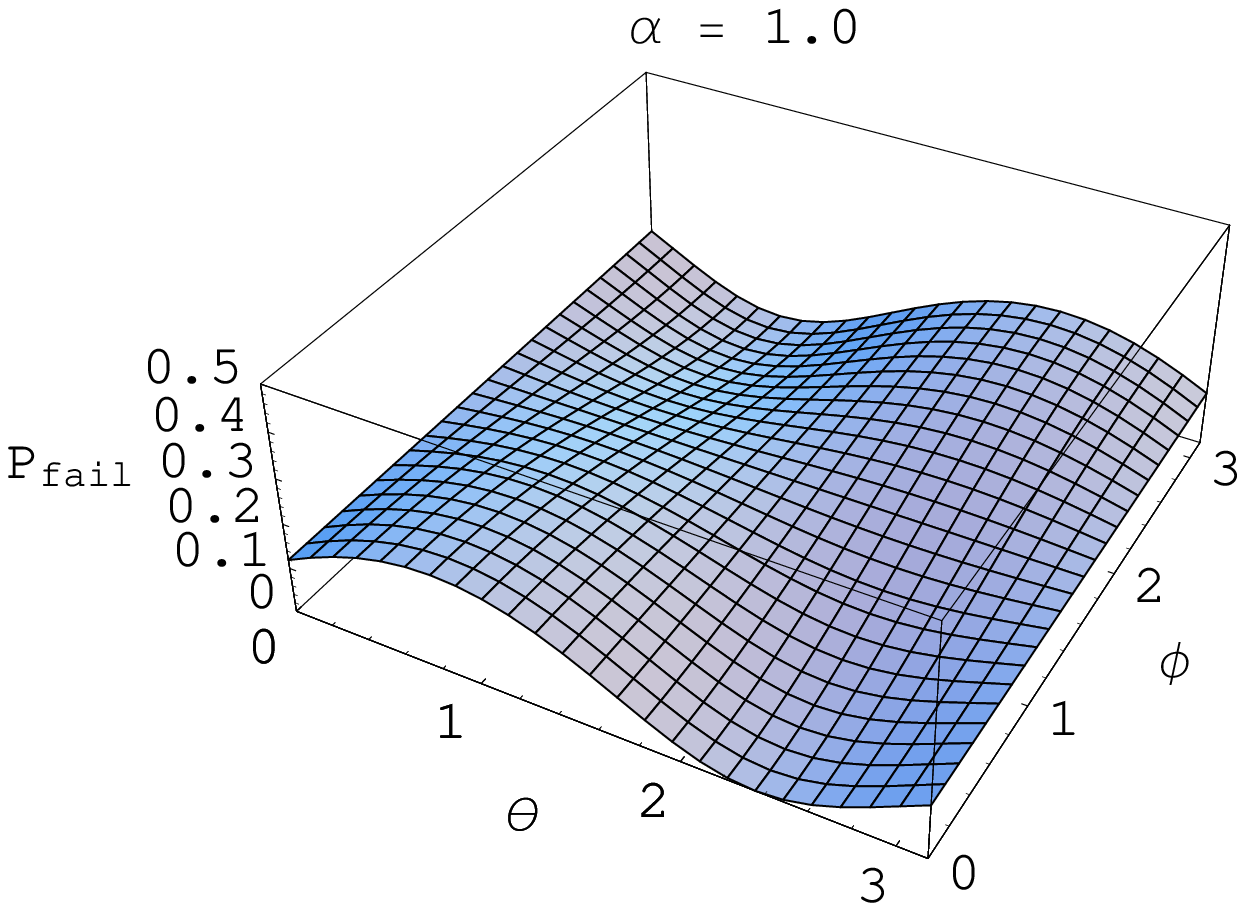}
(c)\includegraphics[width=0.3\textwidth]{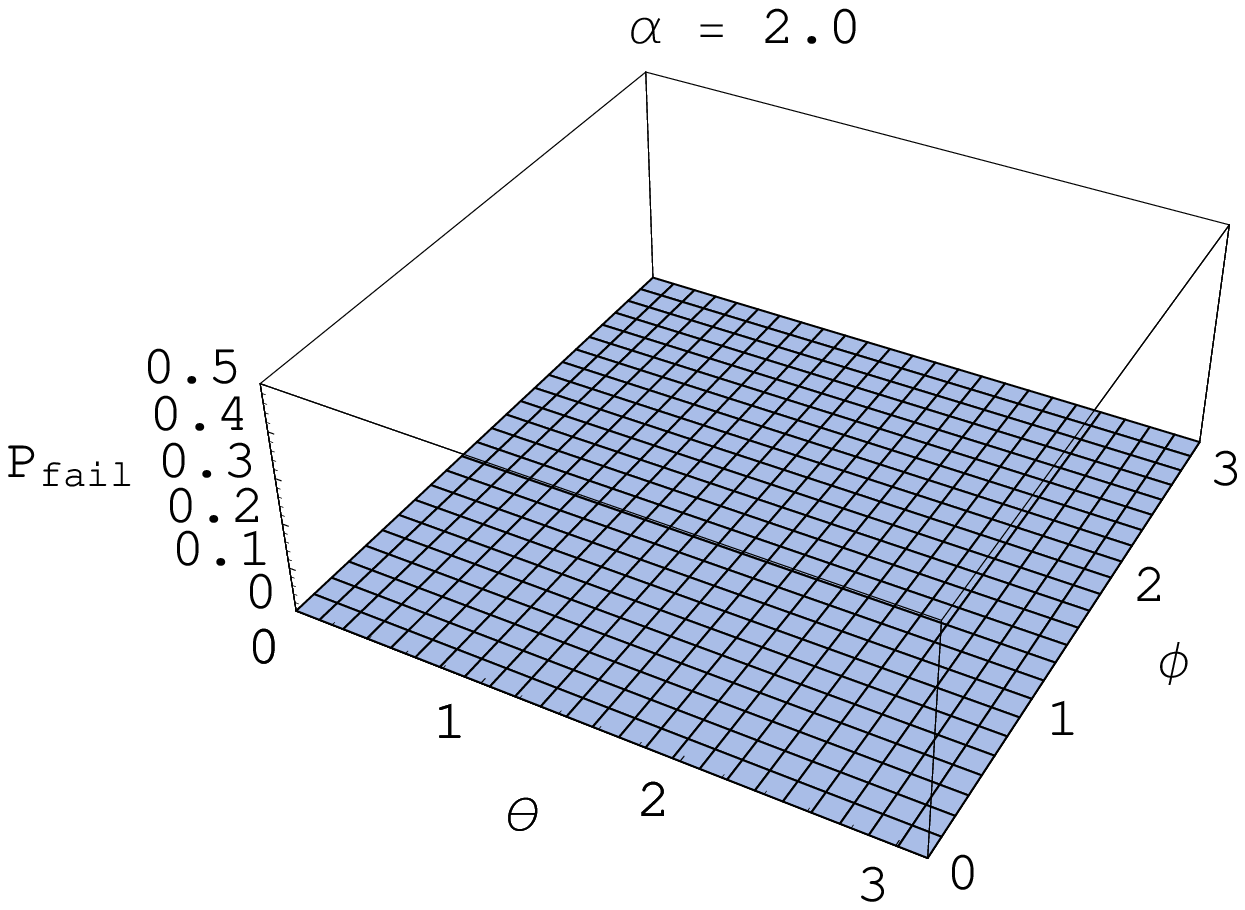}
\caption{The probability of failure of the teleportation protocol ($P_{fail}$) described in the text as per equation~\ref{prob-fail} for $\alpha = 0.5$ (a), $\alpha = 1$ (b) and $\alpha = 2$ (c).  The input state is of the form $\mu = \cos \theta$ and $\nu = e^{i \phi} \sin \theta$.  Note that $P_{fail} < \frac{1}{2}$.}
\label{pfail_plots}
\end{figure}

To estimate the overall probability of success of concatenated teleportations one can sum over the probabilities of events that lead to successful teleportation.  Here we will consider the case where the coherent state Bell state is exact.  As shown above, results of the form $\ket{0,odd}$ require no correction.  It can be shown that results of the form $\ket{odd,0}$ require a $X$ correction which we will assume can be implemented by flipping the reference phase.  As shown in~\cite{vanenk:catteleportation} the total probability of these results ($P_{odd}$) is $\frac{1}{2}$ independent of $\alpha$.  Note that this probability is the maximum probability of success for teleportation using single photon encodings.  The results $\ket{0,even}$ and $\ket{even,0}$ require a $\hat{Z}$ and $\hat{X}\hat{Z}$ corrections respectively.  If the even number is zero then the two cannot be distinguished and the input state cannot be recovered.  This happens with probability $P_{fail}$ which can be shown to be
\begin{equation}
\left| \frac{e^{-\alpha^2}\sqrt{2-2e^{-2 \alpha^2}}(\mu + \nu)}{\sqrt{(2-2 e^{-4\alpha^2})(|\mu|^2 + |\nu|^2 + 2 e^{-2 \alpha^2} Re\{ \nu \mu^*\}})} \right|^2
\label{prob-fail}
\end{equation}
which can be shown to be less than or equal to $\frac{1}{2}$.  Example plots of this probability over all possible inputs states for $\alpha = 0.5, 1$ and $2$ are shown in figure~\ref{pfail_plots}.  The probability that remains must be attributed to the two even results in the Bell state measurement which now can be distinguished.  We will call the probability of obtaining an even result $P_{even}$ which must be $\frac{1}{2} - P_{fail}$ by the argument just made.  

\begin{figure}
\begin{center}
\includegraphics[width=0.45\textwidth]{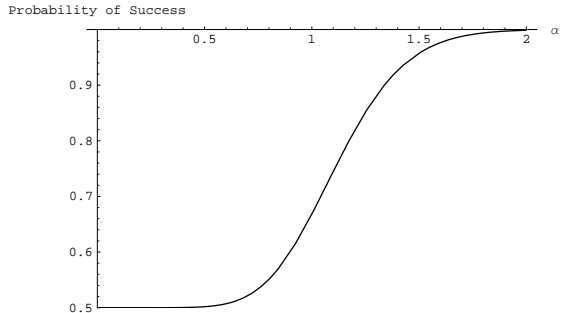} 
\end{center}
\caption{The minimum success probability ($P_{succ}$) over all input states as a function of $\alpha$.}
\label{teleport-min-prob}
\end{figure}

The probability that an ``easy'' correction (i.e. the identity or an $\hat{X}$) need be applied to teleportation is $P_{odd}$.  If however one obtains an even count with probability $P_{even}$ then all is not lost.  When this state is teleported again, if an even result is obtained again with probability $P_{even}$ then the output is an easy correction away from performing teleportation.  This is a result of $\hat{Z}^2$ being the identity. The total probability of getting two consecutive even results is $P_{even}^2$.  However, one may have obtained an odd result before finally getting the second even.  This will occur with probability $P_{even}P_{odd}P_{even}$.  So if teleportation can be performed repeatedly, then the probability of obtaining successful teleportation which is an easy correction away from the input state will approach
\begin{equation}
P_{succ} = P_{odd} + \sum_{n=0}^\infty P_{even} P_{odd}^n P_{even}.
\label{first probability sum}
\end{equation}
Since $P_{odd} < 1$, the sum evaluates to
\begin{equation}
P_{succ} = P_{odd} + \frac{P_{even}^2}{1-P_{odd}}.
\end{equation}
Since $P_{odd} = \frac{1}{2}$ and $P_{even} = \frac{1}{2} - P_{fail}$ this expression is
\begin{equation}
P_{succ} = 1 - 2(P_{fail} - P_{fail}^2).
\end{equation}
If only a set maximum number of teleportations are allowed then equation~\ref{first probability sum} is reduced by removing positive quantities from the sum.  Hence $P_{succ}$ is the maximum probability of success using this method.  From the note above $P_{fail}$ is less than $\frac{1}{2}$ for all values $\alpha$.  This means that $P_{succ}$ is greater than or equal to $\frac{1}{2}$.  
As $P_{succ}$ varies over all inputs states it is minimised for some particular input state given a particular $\alpha$.  This minimum can be traced as $\alpha$ increases as is done in figure~\ref{teleport-min-prob}.  
This shows that in principle this method is capable of qubit teleportation using linear optics and photo detection with a probability greater than $\frac{1}{2}$ for all $\alpha$.  When $\alpha = 1$ the minimum of $P_{succ}$ is $0.67$ which is indicative of the middle ground nature of coherent state encoding at this level. 

\subsection{Teleportation with the squeezed single photon}

In order to perform teleportation in CSQC a source of odd coherent state superpositions is required.  Here we will analyze the response of the teleportation fidelity when the squeezed single photon is used where an odd coherent state superposition is required.  The teleportation fidelity is expected to decrease from unity because the fidelity of the odd coherent state superpotion with the squeezed single photon is less than unity.  We will also consider the effects of losses in the photon counting detectors and will show that this loss dominates any decrease in fidelity as opposed to the decrease from using the squeezed single photon.

\subsubsection{Squeezed single photon}

\begin{figure}
\begin{center}
(a)\includegraphics[width=0.45\textwidth]{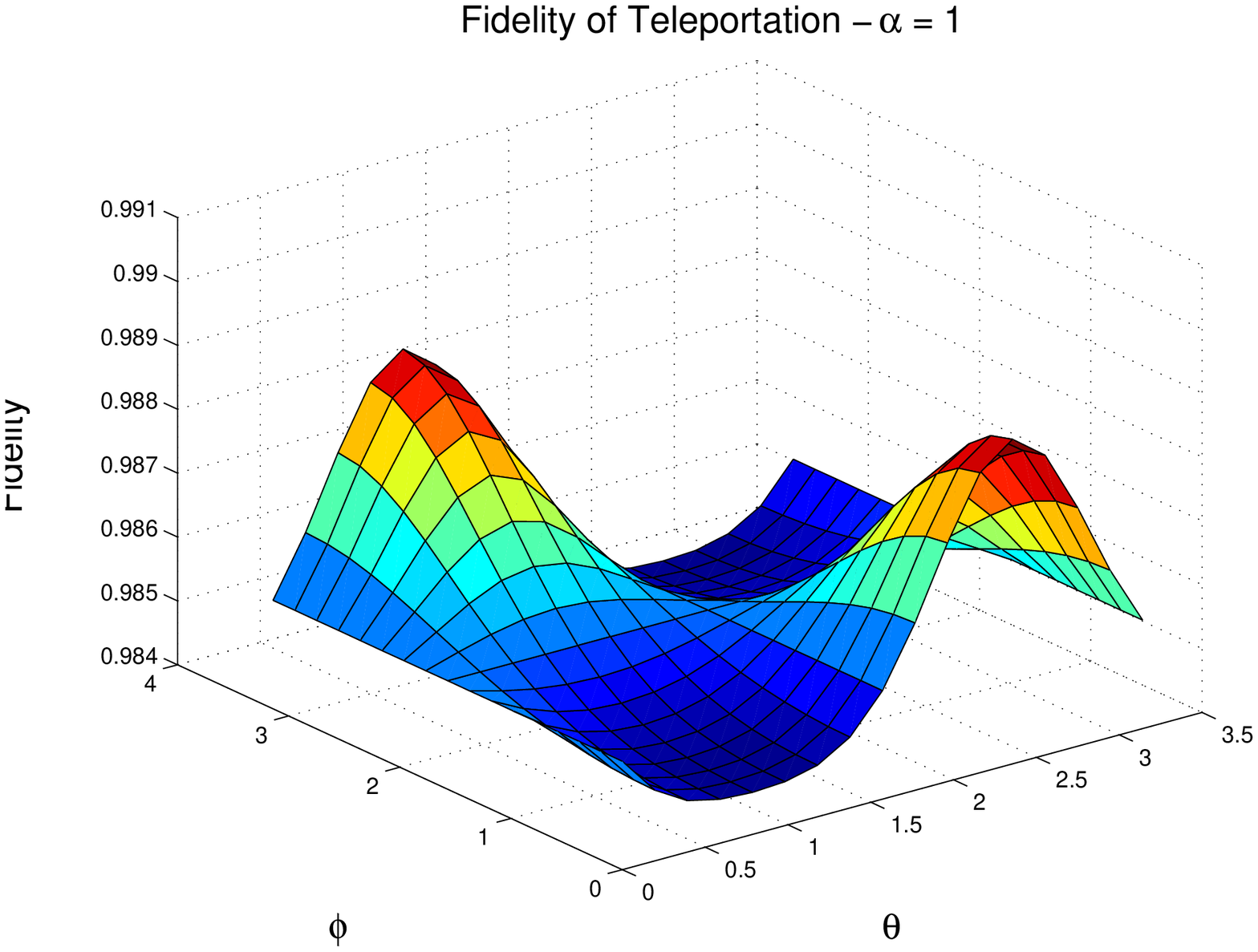} \\
(b)\includegraphics[width=0.45\textwidth]{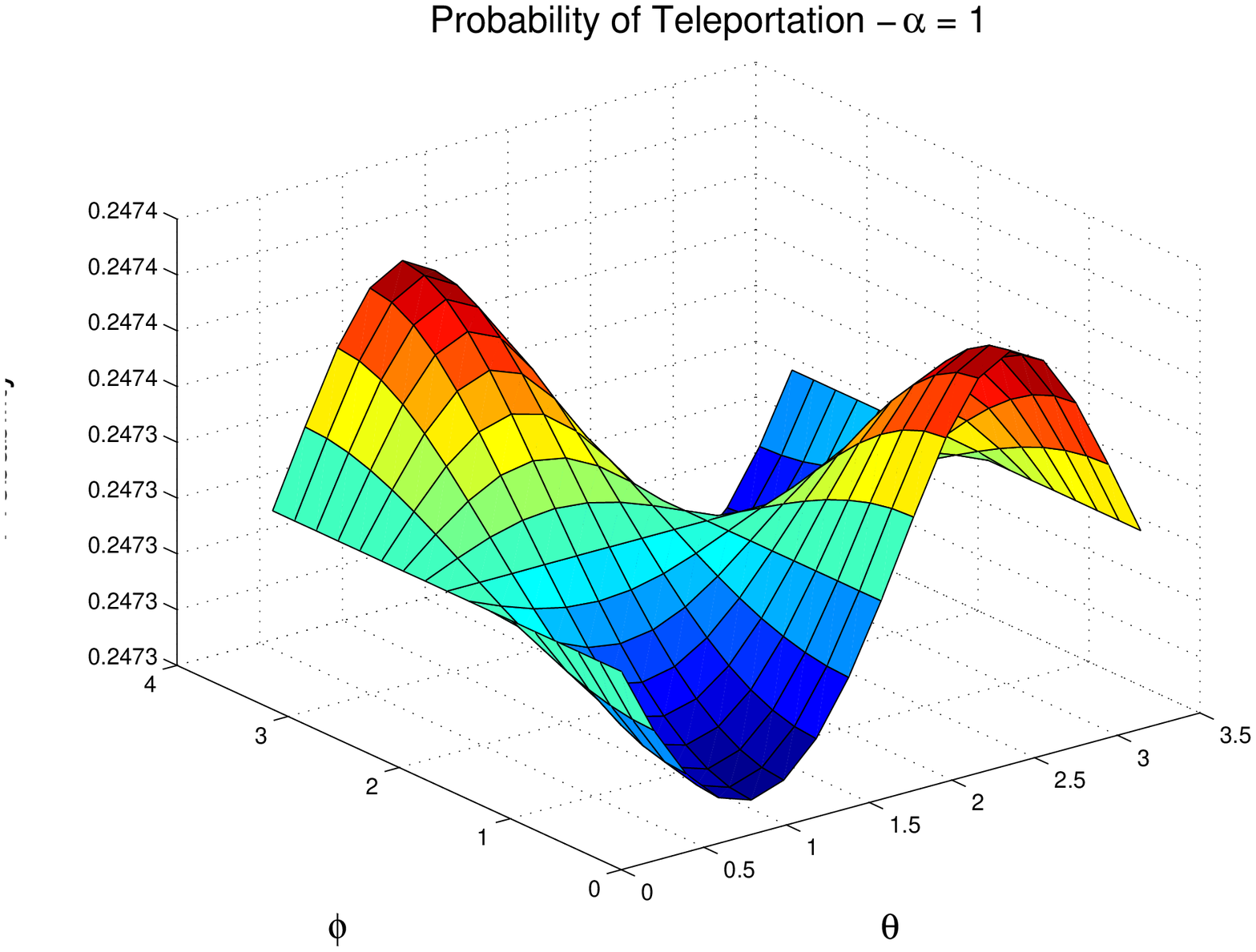}
\end{center}
\caption{A plot of the teleportation fidelity (a) and probability (b) as a function of the input state.  The angles of the input state are defined in equation~\ref{instate}.  The angles $\theta$ and $\phi$ are actually periodic in $2 \pi$ but only half of this region is shown as the remainder is just the mirror image of the plot when continued.  The size of the coherent state qubits is $\alpha = 1$ which means the entanglement is created by a cat state of size $\alpha = \sqrt{2}$.}
\label{tele-fid}
\end{figure}

Figure~\ref{tele-fid} shows the results of a numerical calculation of the fidelity and probability of teleportation when a split squeezed single photon is used as the entanglement resource over a range of possible input states.  
The input states are defined by the two angles $\theta$ and $\phi$ as
\begin{equation}
\label{instate}
\ket{\phi_{in}} = \cos \theta \ket{\alpha} + e^{i \phi} \sin \theta \ket{-\alpha}.
\end{equation}
The fidelity of teleportation for a given input state is the overlap of the input state and the output state squared.  The plot on the right on this figure shows the probability of obtaining the photon number detection result that results in the state which requires no corrections.  The other three bell state measurements could possibly be accepted but $X$ and $Z$ corrections must be applied.  

\subsubsection{Introduction of loss}

\begin{figure}
\begin{center}
(a)\includegraphics[width=0.45\textwidth]{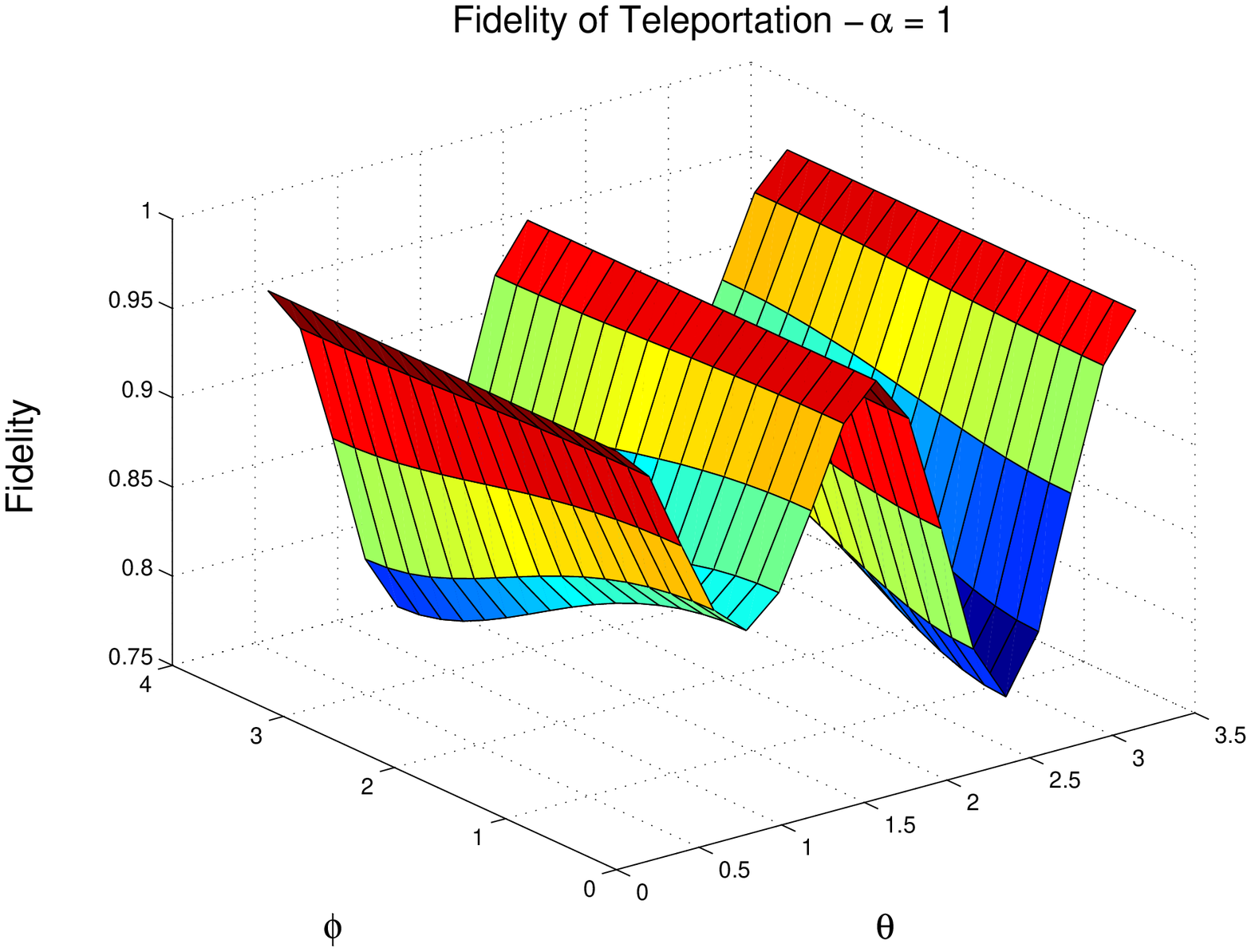} \\
(b)\includegraphics[width=0.45\textwidth]{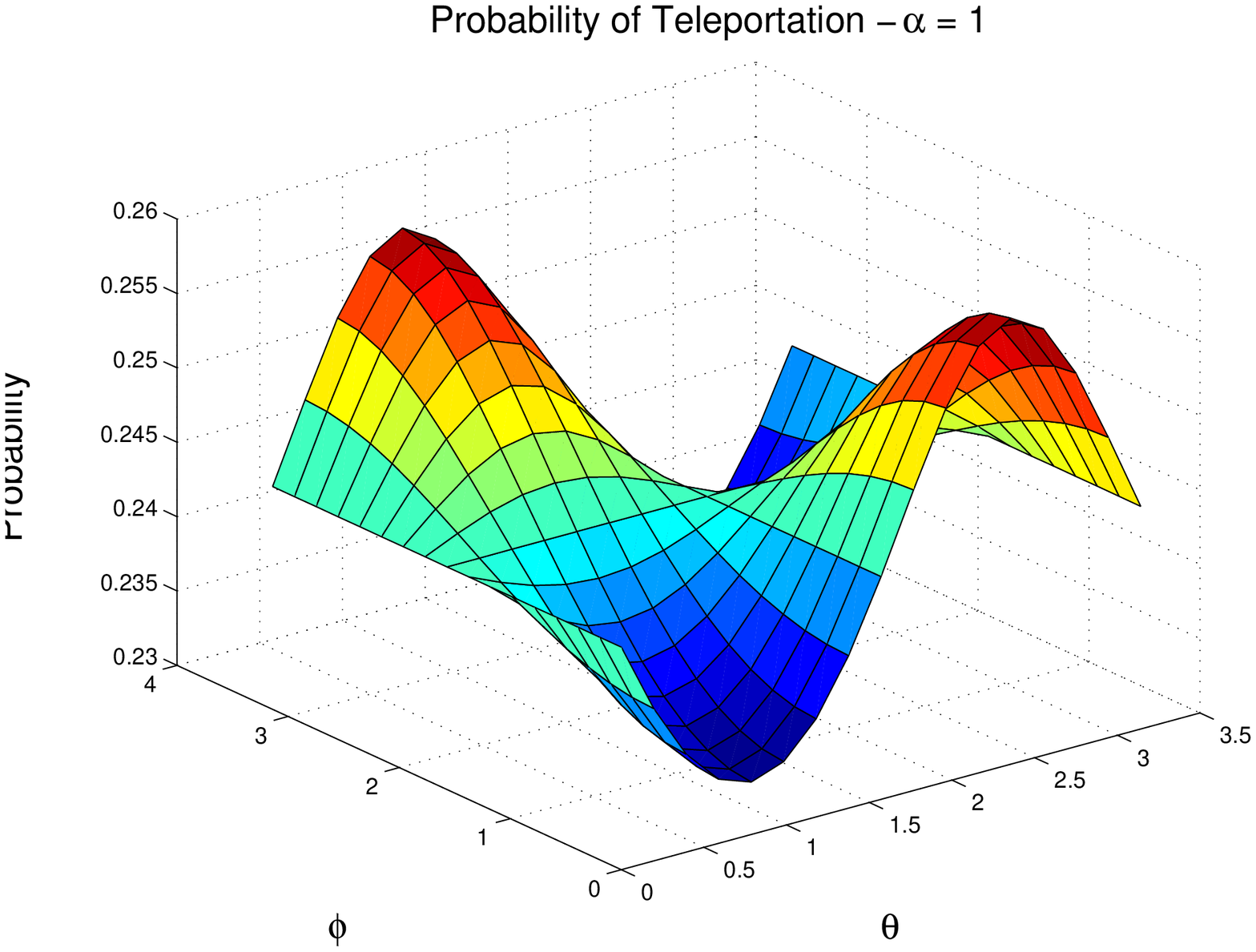}
\end{center}
\caption{This figure shows plots of the same style as figure~\ref{tele-fid} but the detectors in this simulation are 90\% efficienct.}
\label{tele-fid-90eff}
\end{figure}

\begin{figure}
\centerline{
\includegraphics[width=0.45\textwidth]{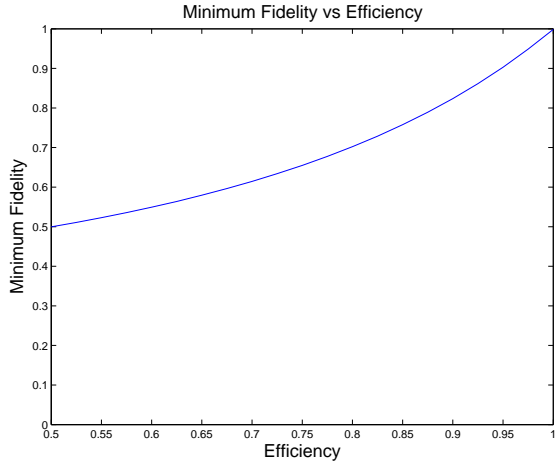}
}
\caption{This figure shows a plot of the minimum fidelity over all possible input states within the space of qubits to the teleporter as a function of detector efficiency.}
\label{multi-min-fid}
\end{figure}

This protocol relies on perfect photon number detection to perform the teleportation.  Here we analyse the effects on teleportation fidelity when the detectors are inefficient while continuing to use the split squeezed single photon as the source of entanglement.  The results shown in figure~\ref{tele-fid} show the fidelity of teleportation as a function of the input state as per equation~\ref{instate} on the left and the probability of performing the teleportation without needing a correction on the right.   Figure~\ref{tele-fid-90eff} shows results in the same format as figure~\ref{tele-fid} but for the case when detection is 90\% efficient.
Note here that the minimum fidelity over all input states has decreased and the probability of a detection result varies much more for the different input states. The increase in probability of success for certain states in the 90\% efficient scenario is an indication of high probability two photon terms in the detection corrupting the measurement.  The output state of the teleporter will be of opposite parity to the desired state when a two photon count is included accidentally.  Hence this high probability corresponds to a drop in the fidelity of the output.
The minimum fidelity over all input states is plotted as a function of efficiency in figure~\ref{multi-min-fid}.

\section{The superposition gate}

A uniquely quantum mechanical effect in quantum computation is being able to move from the qubit basis states into a superposition of these basis states.  An example of a gate which performs an operation of this kind is the Hadamard gate.  The Hadamard gate in the coherent state encoding is written as 
\begin{eqnarray*}
\ket{\alpha} & \rightarrow &  \sqrt{\frac{1}{2}}(\ket{\alpha} + \ket{-\alpha}) \\
\ket{-\alpha} & \rightarrow &  \sqrt{\frac{1}{2}}(\ket{\alpha} - \ket{-\alpha}).
\end{eqnarray*}
This transformation is non-unitary as it takes non-orthogonal states to orthogonal states.  To implement this gate we use the methods described in~\cite{ralph:catcomputing}.  This requires moving outside of the qubit space (i.e. the space spanned by superpositions of $\ket{\alpha}$ and $\ket{-\alpha}$) then projecting back to achieve required phase factors.  The projection onto this subspace is achieved by teleporting the displaced state.  

\subsection{Gate specification}

The procedure to create the Hadamard gate using the coherent state encoding proceeds as follows.  Writing out a general state with this entanglement as per equation~\ref{ent_sys} and then applying a control sign (CSIGN) gate leads to the state
\begin{widetext}
\begin{equation}
\mu \left( \ket{\alpha, \alpha, \alpha} - \ket{\alpha,-\alpha,-\alpha} \right) + \nu \left( \ket{-\alpha,\alpha,\alpha} + \ket{-\alpha,-\alpha,-\alpha} \right).
\end{equation}
\end{widetext}
Projecting the first and second modes onto the odd parity cat state results in the state
\begin{equation}
\mu \left( \ket{\alpha} + \ket{-\alpha} \right) + \nu \left( -\ket{\alpha} + \ket{-\alpha} \right).
\end{equation}
Applying a bit flip gate or $\hat{X}$-gate leads to 
\begin{equation}
\mu \left( \ket{\alpha} + \ket{-\alpha} \right) + \nu \left( \ket{\alpha} - \ket{-\alpha} \right)
\end{equation}
which is the Hadamard transformation.  Reference~\cite{ralph:catcomputing} shows a way in which to build a CSIGN gate using this encoding, however it is assumed that the coherent states are well seperated.  The CSIGN gate is a symmetric beamsplitter with a reflectivity chosen so that the coherent states displace each other in such a way that projecting back onto the cat state basis results in a sign change for the appropriate state.  This does not apply directly to the regime of small cat state to which the squeezed single photons are good approximations.  However, this scheme can be used as a guide on how to construct a gate that may apply for small cat state given certain restrictions.  

Starting from equation~\ref{ent_sys} and applying a symmetric beamsplitter to the first two modes leaves the state as
\begin{widetext}
\begin{equation}
\mu \left( \ket{\alpha e^{i \theta}, \alpha e^{i \theta}, \alpha} - \ket{\alpha e^{-i \theta},-\alpha e^{-i \theta},-\alpha} \right) + \nu \left( \ket{-\alpha e^{-i \theta},\alpha e^{-i \theta},\alpha} - \ket{-\alpha e^{i \theta},-\alpha e^{i \theta},-\alpha} \right).
\end{equation}
\end{widetext}
Now one should perform a projection onto the odd cat state in modes one and two.  As shown in~\cite{ralph:catcomputing} provided the displacements are not too large one can perform photon counting with only small errors. As a special case of this if only the one photon term is accepted then the state transforms to
\begin{widetext}
\begin{equation}
e^{-|\alpha|^2} \alpha^2 \mu \left( e^{i 2\theta} \ket{\alpha} + e^{-i 2\theta} \ket{-\alpha} \right) - e^{-|\alpha|^2} \alpha^2 \nu \left(  e^{-i 2\theta} \ket{\alpha} + e^{i 2\theta} \ket{-\alpha}\right).
\end{equation}
\end{widetext}
Plugging in $\theta = \pi/8$, then up to a global phase factor and ignoring the normalisation the transformation can be written
\begin{equation}
\label{hadabar}
\mu \ket{\alpha} + \nu \ket{-\alpha} \rightarrow (\mu + i \nu) \ket{\alpha} - (i \mu + \nu) \ket{-\alpha}.
\end{equation}
This transformation is equivalent to a Hadamard transformation provided one can perform $\hat{Z}$ operations.   That is if the transformation 
\begin{equation}
\mu \ket{\alpha} + \nu \ket{-\alpha} \rightarrow \mu \ket{\alpha} - i \nu \ket{-\alpha}
\end{equation}
is applied before and after the transformation in equation~\ref{hadabar} then a Hadamard transformation is obtained. The transformation in equation~\ref{hadabar} is not the Hadamard operation but is still very useful as it takes qubit basis states into superpositions of both qubit basis states. We will call the transformation in equation~\ref{hadabar} the ``\emph{rotated Hadamard}'' transformation which we will denote as $\hat{\bar{H}}$.  Note that this transformation is exact for any size of coherent states as long as the photon number measurement obtains two single photon counts. 

\subsection{Probability of success}

The probability of obtaining a photon count of $m$ photons in one and $n$ in the other (i.e. projecting onto the state $\ket{n,m}$) is
\begin{widetext}
\begin{equation}
\label{hadab-pro}
\left( \frac{1-2 \mathrm{Re}\left\{ \mu^* \nu e^{-2|\alpha|^2} \right\}}{(1+2 \mathrm{Re}\left\{ \mu^* \nu e^{-2|\alpha|^2} \right\})(1-e^{-4|\alpha|^2})} \right) \left( \frac{\alpha^{2(n+m)} e^{-2|\alpha|^2}}{n! m!} \right).
\end{equation}
\end{widetext}
The important term here is the one on the right involving $n$ and $m$.  The probability falls as the factorial of the photon number but also as $\alpha^2$ to the power of the sum of the two photon numbers.  So for $\alpha < 1$ there is a major advantage when losses are considered as the probability of higher photon number terms already reduce quickly as well as having the advantage of the detector efficiency to reject error counts. 

\section{Combining Gates}

\subsection{The candidate computation}

\begin{figure}
\setlength{\unitlength}{1cm}
\centerline{
\begin{picture}(9,2)
\put(0,1){\line(1,0){1}}
\put(1,0){\framebox(2,2){$\hat{H}$}}
\put(3,1){\line(1,0){1.25}}
\put(4,0.5){\line(1,2){0.5}}
\put(5,0.5){\line(-1,2){0.5}}
\put(4,0.5){\line(1,0){1}}
\put(4,0.5){\makebox(1,1){$\phi$}}
\put(4.75,1){\line(1,0){1.25}}
\put(6,0){\framebox(2,2){$\hat{H}$}}
\put(8,1){\vector(1,0){1}}
\end{picture}
}
\caption{A schematic drawing of a simple, non-trivial experiment involving a qubit.  The input state can be any qubit but will be set to the state $\ket{1}$ throughout.  The boxes represent the Hadamard gates and the triangle represents a phase shift between the qubits.  The output is detected in the computational basis.}
\label{fringe-setup}
\end{figure}
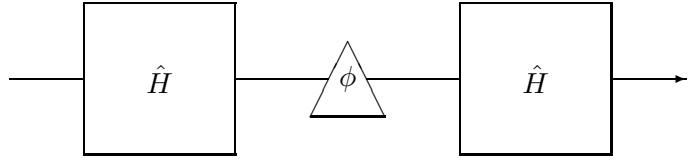

One of the simplest non-trivial computation that can be performed with a qubit is two Hadamard gates with a phase shift between them.  This arrangement is shown in figure~\ref{fringe-setup}.
Detection in the computational basis at the end of this experiment will reveal different probabilities when the phase shift between the two Hadamard gates is changed.  The variation of probability with respect to the phase shift is a uniquely quantum mechanical property, hence verifying the existence of a quantum bit.  This kind of experiment is equivalent to an interferometer where the path length between the two arms can be varied.  Hence a plot of the probability of detecting one of the basis states with respect to the phase shift is called a fringe.  With perfect qubits the period of the fringe should be $\pi$ and the visibility should be unity as the probability should drop to zero for some phase.  The visibility for this fringe is defined here as
\begin{equation}
V = \frac{P_{max} - P_{min}}{P_{max} + P_{min}}.
\end{equation}
Since entanglement generated by a squeezed photon state is not exactly a cat state then the visibility is expected to be slightly less than unity. 

\subsection{Implementing the computation using CSQC}

The exact Hadamard transformation is not available when considering the small cat state generated by the squeezed single photon state.  However, the $\hat{\bar{H}}$ gate as described previously is avaliable.  In this numerical calculation, the first Hadamard gate has the effect of preparing the state $\ket{0} + \ket{1}$.  Using the squeezed single photon states the state $\ket{\alpha} - \ket{-\alpha}$ can be generated so this state will be used instead.  This removes the necessity for the first Hadamard gate.  The phase shift can be implemented by displacing the cat state in the imaginary direction then projecting back into the computational basis.  It is hoped that the detection in the $\hat{\bar{H}}$ gate will perform this projection when the appropriate measurement results is achieved.  After this displacement the $\hat{\bar{H}}$ gate is applied. 
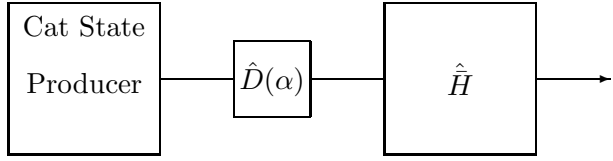
\begin{figure}
\setlength{\unitlength}{1cm}
\centerline{
\begin{picture}(8,2)
\put(0,0){\framebox(2,2){}}
\put(0,1){\parbox{2\unitlength}{\begin{center}Cat State Producer\end{center}}}
\put(2,1){\line(1,0){1}}
\put(3,0.5){\framebox(1,1){$\hat{D}(\alpha)$}}
\put(4,1){\line(1,0){1}}
\put(5,0){\framebox(2,2){$\hat{\bar{H}}$}}
\put(7,1){\vector(1,0){1}}
\end{picture}
}
\caption{An experiment in the same spirit as the one in figure~\ref{fringe-setup}.  The cat state producer is the squeezed single photon state, the phase shift is now a displacement and the Hadamard is the rotated Hadamard gate.}
\label{fringe-actual}
\end{figure}
With an ideal odd cat state and an ideal phase shift the input state to the rotated Hadamard is 
\begin{equation}
\ket{\alpha} - e^{i\theta} \ket{-\alpha}
\end{equation}
ignoring normalisation.  After applying the $\hat{\bar{H}}$ gate the state transforms to
\begin{equation}
(1-i e^{i\theta}) \ket{\alpha} - (i - e^{i \theta}) \ket{-\alpha}.
\end{equation}
When this state is measured in the coherent state basis it should give a sinusoidal probability response as the displacement is changed.  Ideally the probability of the two coherent states should be equal when $\theta = 0$.  

The measurement in the computational basis can be performed by combining the signal which is a superposition of $\ket{\alpha}$ and $\ket{-\alpha}$ with another signal prepared in the state $\ket{\alpha}$ on a 50:50 beamsplitter. 
The effect on either coherent state is
\begin{eqnarray*}
\ket{\alpha} \ket{\alpha} & \rightarrow & \ket{0}\ket{\sqrt{2} \alpha} \\
\ket{-\alpha} \ket{\alpha} & \rightarrow & \ket{-\sqrt{2} \alpha} \ket{0}.
\end{eqnarray*}
The two coherent states can now be distinguished by a measurement on the two modes.  If there are no photons in one mode and one or more in the other mode then the detection has succeeded and by the mode in which non-zero photon number occurred the sign of the coherent amplitude can be determined.  The measurement fails if zero photons are detected in both modes.  This occurs with a probability of $e^{-|\alpha|^2}$ and will approach zero quickly as $\alpha$ grows.  Also the detection need not require efficient detection.  If a photon is lost, the probability of the detection drops but no errors will occur.

\begin{figure}
\begin{center}
(a)\includegraphics[width=0.45\textwidth]{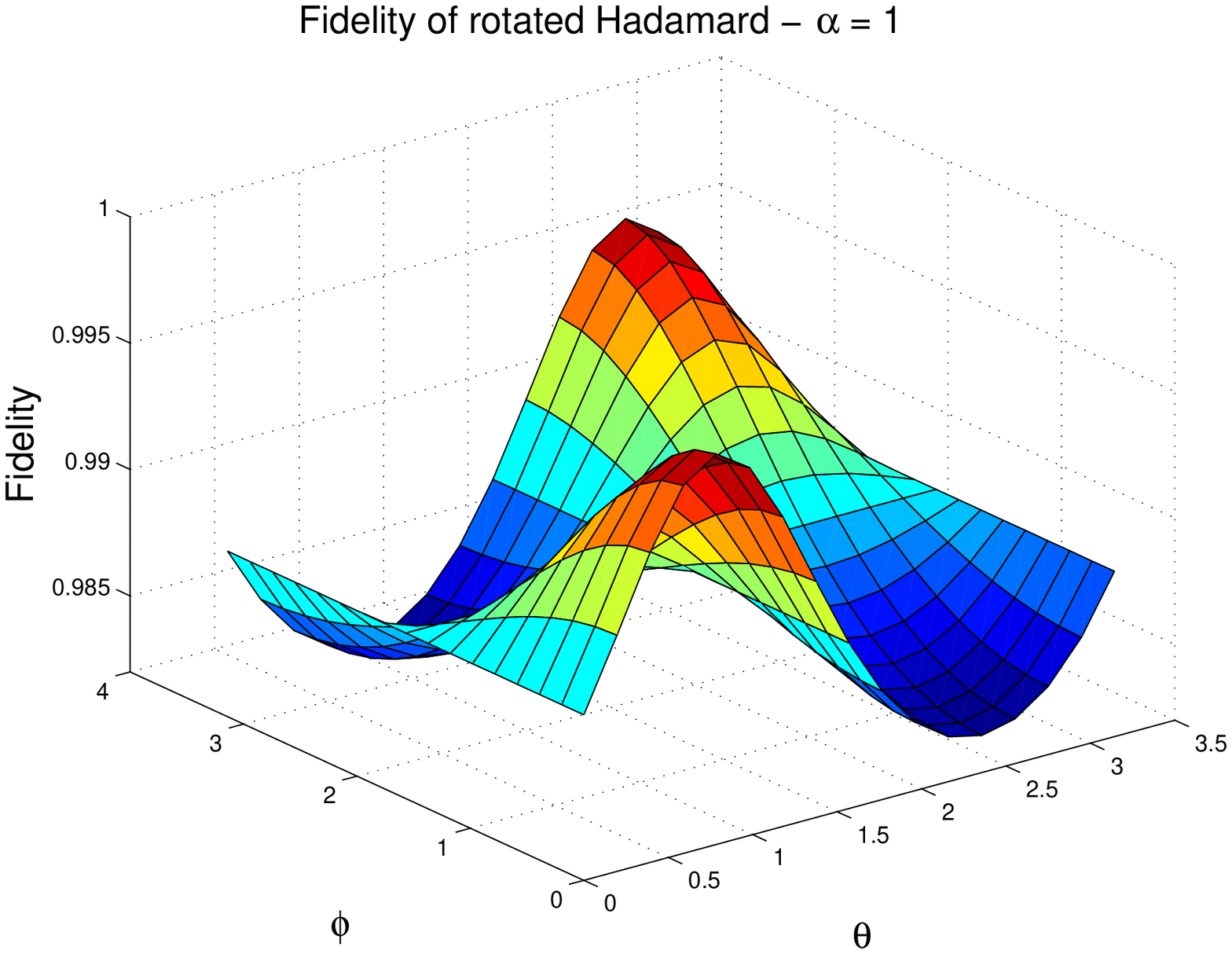} \\
(b)\includegraphics[width=0.45\textwidth]{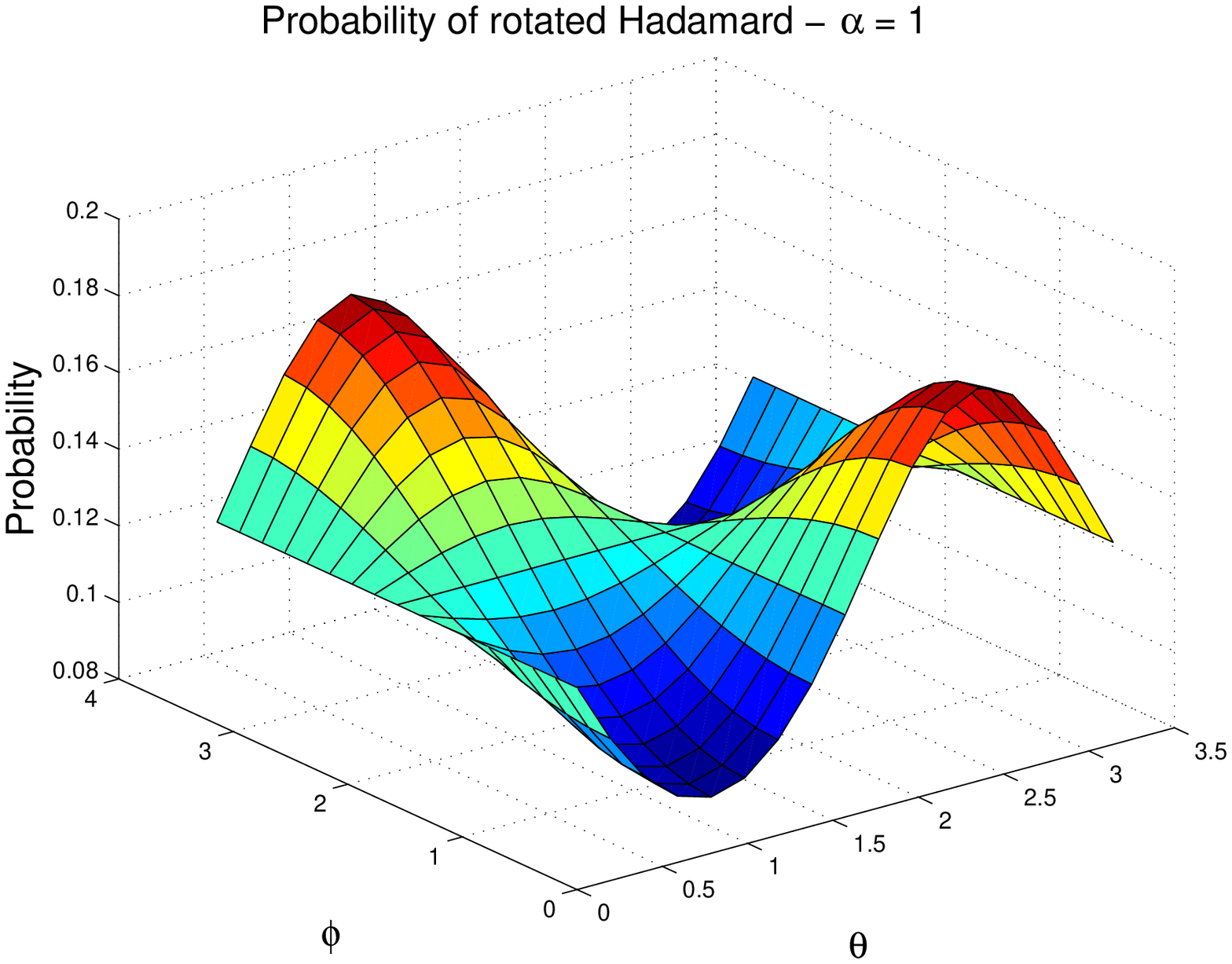}
\end{center}
\caption{The plots in this figure are in the same style as figure~\ref{tele-fid} but are for the $\hat{\bar{H}}$ gate with the coherent amplitude of the qubits at $\alpha = 1$.  Note that for such small coherent state qubits the squeezed single photon is very close to a cat state.  This means that the fidelity should be close to one.}
\label{hadab-sim}
\end{figure}

The performance of this gate using the squeezed single photon states as cat states can be analysied in a way similar to the teleportation gate.  Figure~\ref{hadab-sim} shows the fidelity of the gate compared to the expected output shown in equation~\ref{hadabar}.
The plot on the right of this figure is the probability of the gate functioning over the range of input states.  This plot agrees well with the prediction of the $(n,m) = (1,1)$ count used in equation~\ref{hadab-pro}.  Note that the probability of a successful detection depends on the input state of the qubit.  One could consider that some sort of measurement has been made on the qubit as different probabilities apply for different input states (see~\cite{lund:comparison}).  However this does not destroy the calculation as the basis qubits have a non-zero overlap.  

\begin{figure}
\begin{center}
(a)\includegraphics[width=0.43\textwidth]{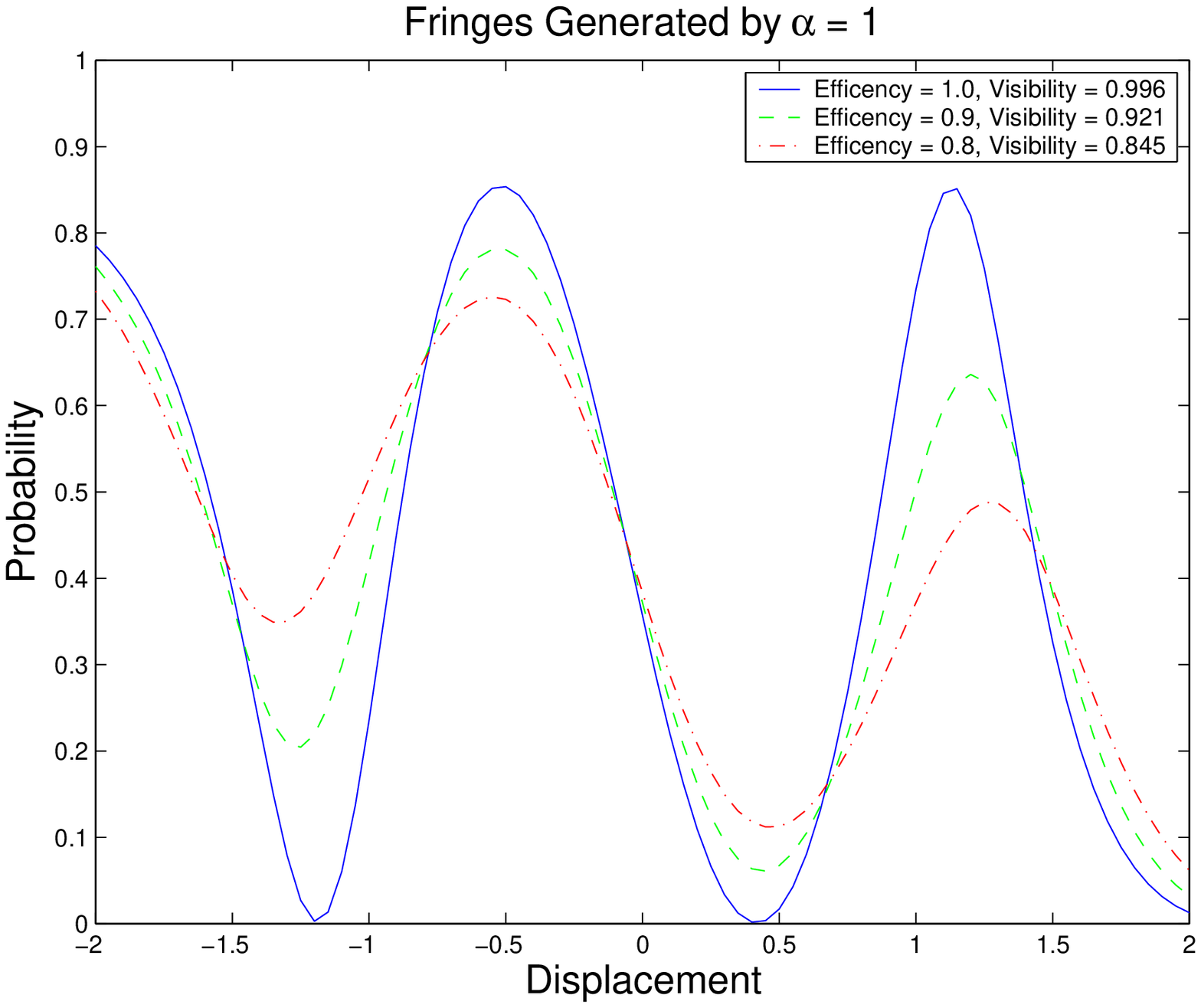} \\
(b)\includegraphics[width=0.43\textwidth]{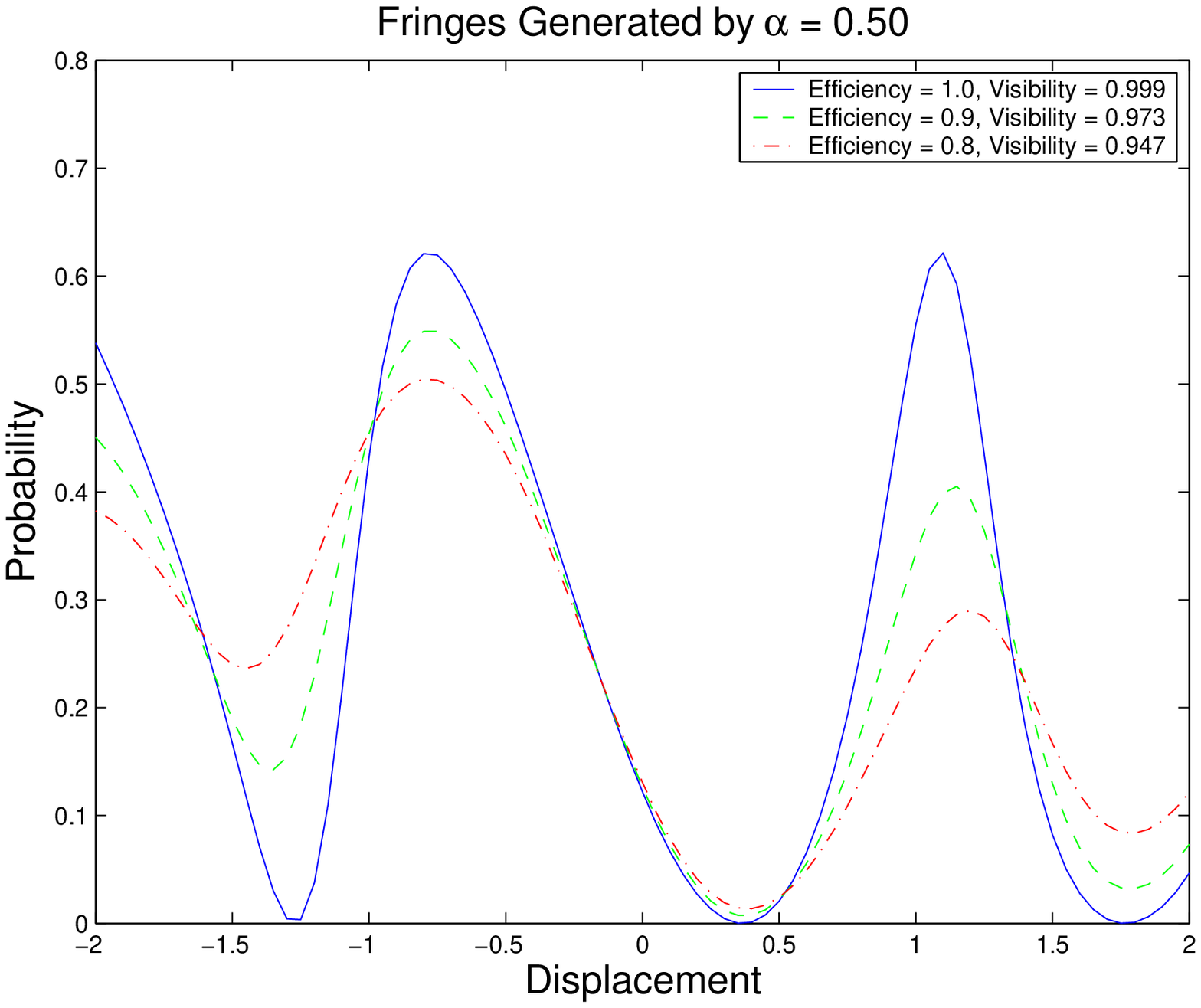} \\
(c)\includegraphics[width=0.43\textwidth]{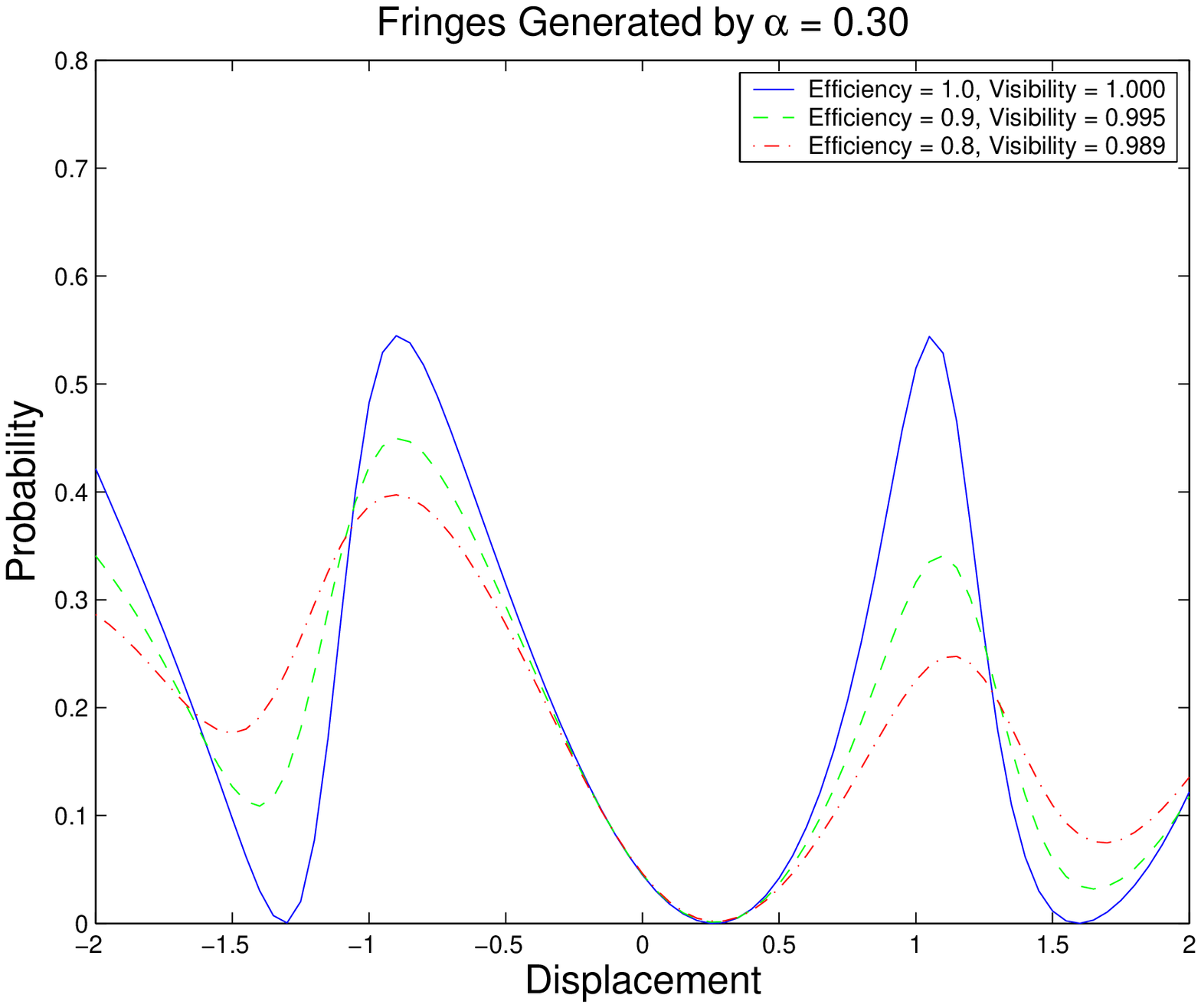}
\end{center}
\caption{Plots of the probability of detecting the coherent state $\alpha$ at the output of the device shown in figure~\ref{fringe-actual}.  The non-sinusoidal nature of this function is due to the reliance of the $\hat{\bar{H}}$ gate to project the displaced cat state back on to the computational basis states.  The plots are for a variety of $\alpha$.  $\alpha = 1$ in (a), $\alpha = 0.5$ in (b) and $\alpha = 0.3$ in (c).}
\label{fringes}
\end{figure}

\begin{figure}
\begin{center}
\includegraphics[width=0.45\textwidth]{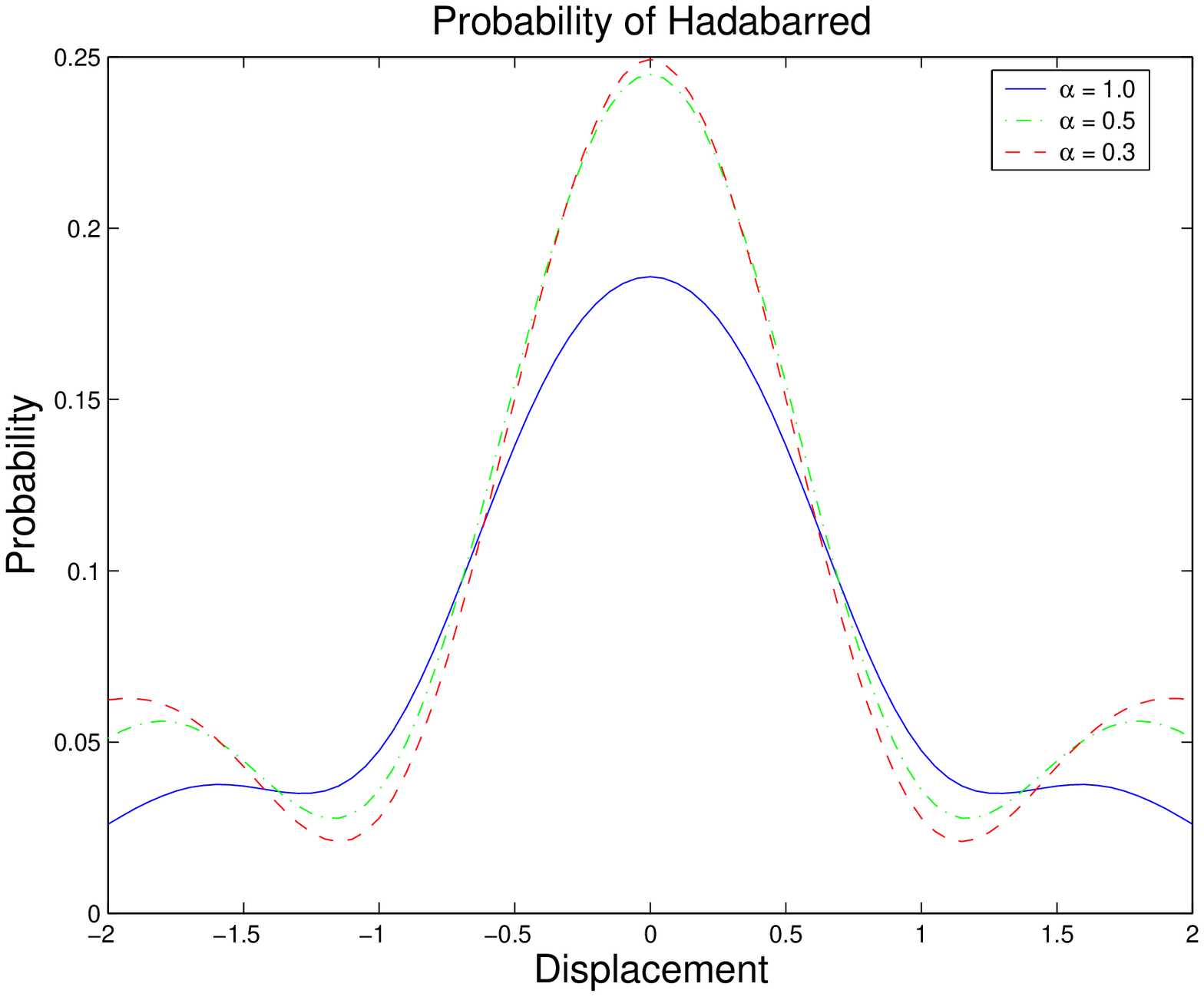}
\end{center}
\caption{This figure shows the probability that the $\hat{\bar{H}}$ gate gives a successful post-selection as a function of the displaced input.}
\label{fringe-hada-prob}
\end{figure}

We also perform the numerical calculation for arrangement depicted in figure~\ref{fringe-actual} which should generate a sinusoidal variation in the probability distribution of one of the basis states at the output.  The fringes which result from this simulation are shown in figure~\ref{fringes}.
This plot shows the probability of obtaining one particular coherent state using the detector involving the mixing of the output state with a coherent state described above. Note that the probability of the $\hat{\bar{H}}$ gate functioning is not included in this probability.  It is shown in figure~\ref{fringe-hada-prob}.
The first thing to note is that the fringes are not sinusoidal in nature.  This is due to the reliance of the $\hat{\bar{H}}$ gate to project the displaced cat state back into the computational basis.  However, the visibilities for each fringe can still be calculated and are shown in the legend of each graph.  The multiple plots on each graph shows how the fringes change as the detector efficiency of the detectors in the $\hat{\bar{H}}$ gate change.  The fringe visibilities remain high for poor detectors due to the large drop off of the photon counting events in the detectors.  

\section{Conclusion}

We have shown in this paper that demonstrations of the basic functionality of quantum computation gates based on coherent state quantum bits is within reach of current technology.  Superpositions of coherent states with relatively small amplitudes can be well approximated by the squeezed single photon state.  Furthermore, gates which use superpositions of coherent states as a resource can utilise the squeezed single photon as this resource and still function with high fidelities.   The small coherent amplitudes require some modification of gate operation, but basic functionality can still be achieved.  For the case of teleportation, an improvement in efficiency over photonic systems can be recognised even at the small amplitudes considered here with success probabilities of 67\% with over 99\% fidelity predicted. This is to be compared with the 50\% success probability achieved with basic photonic systems. 
\bibliography{thebib.bib}

\end{document}